\documentclass[12pt,a4paper,dvipdfmx]{article}
\usepackage{graphicx}
\usepackage{amsmath}
\usepackage{url}
\usepackage{bm}
\usepackage{float}

\newcommand\addnum[1]{$^{#1}$}

\begin{document}

\title{\Large\bf
Direct Evidence for Synchronization in Japanese Business Cycle
}

\author{\sl 
  Yuichi Ikeda \addnum1, 
  Hideaki Aoyama \addnum2,\\
  Hiroshi Iyetomi \addnum3,
  Hiroshi Yoshikawa \addnum3
  \\[10pt]
  {\sl\small\addnum1
    Graduate School of Advanced Integrated Studies in Human Survivability,}\\
  {\sl\small 
    Kyoto University, Kyoto 606-8501, Japan}\\
  {\sl\small\addnum2
    Department of Physics, Kyoto University, Kyoto 606-8502, Japan}\\
  {\sl\small\addnum3
    Faculty of Economics, University of Tokyo, Tokyo 113-0033, Japan}
}

\date{}
\maketitle

\begin{abstract}
We have analyzed the Indices of Industrial Production (Seasonal Adjustment Index) for a long period of 240 months (January 1988 to December 2007) to develop a deeper understanding of the economic shocks. 
The angular frequencies estimated using the Hilbert transformation, are almost identical for the 16 industrial sectors.
Moreover, the partial phase locking was observed for the 16 sectors.
These are the direct evidence of the synchronization in the Japanese business cycle.
We also showed that the information of the economic shock is carried by the phase time-series. The common shock and individual shocks are separated using phase time-series. The former dominates the economic shock in all of 1992, 1998 and 2001.
The obtained results suggest that the business cycle may be described as a dynamics of the coupled limit-cycle oscillators exposed to the common shocks and random individual shocks.
\end{abstract}

\bigskip



\bigskip

\noindent\hrulefill

\noindent
{\bf Correspondence\/}:
  Yuichi Ikeda,
  email: ikeda.yuichi.2w@kyoto-u.ac.jp\\[20pt]


\newpage
\section{Introduction}\label{sec:intro}

Business cycles has been defined \cite{Mitchell1964} as a type of fluctuation found in the aggregate economic activity of nations that organize their work mainly in business enterprises: a cycle consists of expansions occurring at about the same time in many economic activities, followed by similar general recessions, contractions, and revivals which merge into the expansion phase of the next cycle; this sequence of change is recurrent but not periodic. Duration wise, one business cycle may vary from more than one year to ten or twelve years, and it is not divisible into shorter cycles of similar character with amplitudes approximating their own.

Business cycle has a long history of theoretical studies \cite{Yoshikawa2000, Aoki2007, Granger1964}.
Samuelson \cite{Samuelson1939} showed that the second-order ordinary linear differential equation based on multiplier and accelerator could generate the cycle in GDP.
Hickes demonstrated that the cycle is sustainable by introducing the ``ceiling'' and the ``floor'' into such model\cite{Hickes1950}.
In contrast, some other theorists paid attention to the non-linearity. Goodwin \cite{Goodwin1951} introduced the nonlinear accelerator in order to generate the sustainable cycle. Kaldor \cite{Kaldor1940} captured the business cycle as a limit cycle.
Subsequently, the idea of non-linearity evolved in the application of the Chaos theory \cite{Lorenz1989, Moseklide1992, Sterman1994}.
It should be noted that the earlier empirical studies have been reported mainly for the US business cycle \cite{Mitchell1964, Tinbergen1939}.

In the neoclassical economics,
predominant causes of the business cycle are considered as real changes in the supply side, such as 
innovation in production technology and shift in fiscal policy instead of nominal demand change due to shift in financial policy.
The most influential theory among the neoclassical economists is the real business cycle (RBC) theory \cite{Kydland1982}, 
whose essential feature is to treat the shock due to technological innovations as the most important cause of the business cycle, as outlined below.

\begin{figure}
\begin{center}
\includegraphics[width=0.4\textwidth]{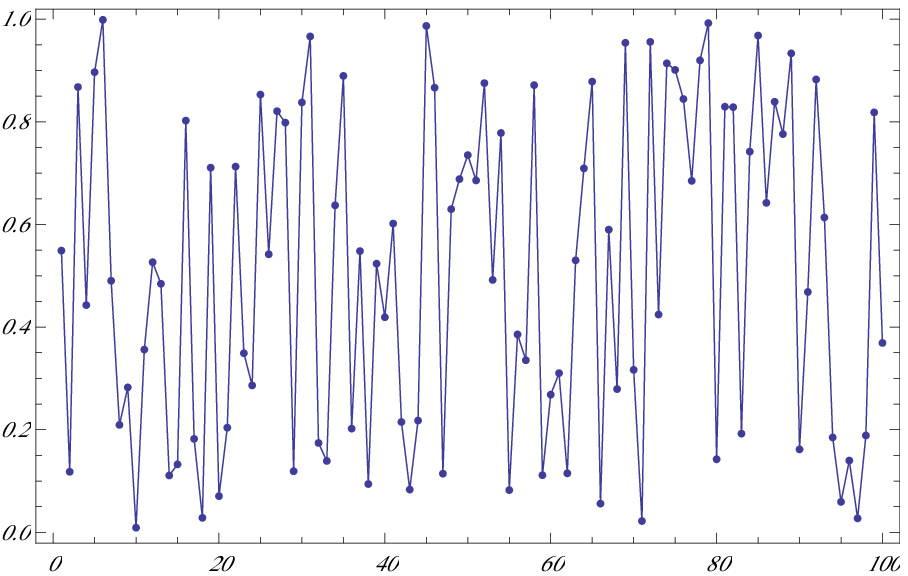}
\includegraphics[width=0.4\textwidth]{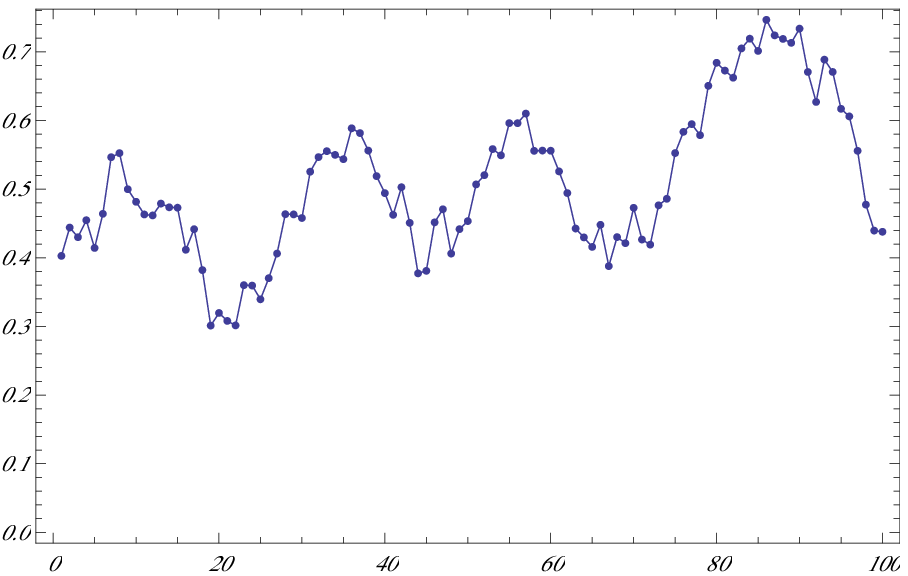}

\caption{
The left panel is uniform random number between 0 and 1 for 100 time points. The right is the moving summation of random numbers for the last 10 time points shown in the left panel.
This demonstrates that the moving summation of random numbers generate cyclical fluctuation.
}
\label{fig:MovSum}
\end{center}
\end{figure}

If we deal with the Cobb-Douglas production function
\begin{equation}
  \label{eq:CobbDouglas}
  Y=AK^\alpha L^{1-\alpha},
\end{equation}
the Solow's residual
\begin{equation}
  \label{eq:Solow}
  \frac{\Delta A}{A} = \frac{\Delta Y}{Y} - \alpha \frac{\Delta K}{K}  -(1-\alpha) \frac{\Delta L}{L}
\end{equation}
is considered as the shock due to technological innovations.
The occurrence of technological innovation could be treated as random process. 
Therefore the Solow's residual is emulated as a moving summation of random shocks, following Slutsky's explanation \cite{Slutsky1937}.
Figure \ref{fig:MovSum} depicts that the moving summation of random numbers generates cyclical fluctuations.
Policy implication of the RBC theory is that the aggregate fluctuation is interpreted as rational or optimal, and any shift in financial policy has no effect on the business cycle.

In this paper, we explored the alternative view for the business cycle in order to develop a deeper understanding of the economic shocks.
Let us assume that two pendulum clocks, which represent industrial sectors, with slightly different oscillation periods are hanged on a beam. The pendulums of these two clocks oscillate with the same period and two clocks show the same hour\cite{Huygens1673}. 
When we move one pendulum clock to another beam, the periods of the oscillation are not same for both the pendulum clocks, and they gradually deviates.
A pendulum clock is considered as a limit cycle oscillator, which receives energy from the rest of the system and releases energy and entropy to the rest of the system during the oscillation. The observed oscillation with the same period is induced by the interaction between two pendulum clocks mediated by the beam. 
This phenomenon is called synchronization, or entrainment \cite{Kuramoto1984, Ikeda2012}.

If we divide the aggregate business cycle into industrial sector, fluctuation of industrial sectors must be synchronized with the same oscillation period. 
Otherwise the aggregate business cycle would not be observed. 
Here, it is noted that the division into industrial sector is regarded as a macroscopic resolution.
The present work is an empirical study of synchronization in the Japanese business cycle using data of the Indices of Industrial Production.

\section{Empirical Analysis}\label{sec:emp}

\subsection{Data}\label{sec:emp:data}

We analyzed data of the Indices of Industrial Production (Seasonal Adjustment Index) for a long period of 240 month (January 1988 to December 2007). 
The database includes the indices of production, shipment and inventory for the following 16 industrial sectors:
steel industry (s1), non-ferrous metal industry (s2), metal products industry (s3), machinery industry (s4), 
transportation equipment industry (s5), precision machinery industry (s6), ceramic products industry (s7), 
chemical industry (s8), petroleum and coal products industry (s9), plastic products industry (s10), 
paper and pulp industry (s11), textile industry (s12), food industry (s13), other industries (s14), 
mining industry (s15), and electric machinery industry (s16). 
The analysis performed in the present study employed just the indices of production for all 16 sectors mentioned above.

\subsection{Stationarization}\label{sec:emp:stat}

\begin{figure}
\begin{center}
\includegraphics[width=0.95\textwidth]{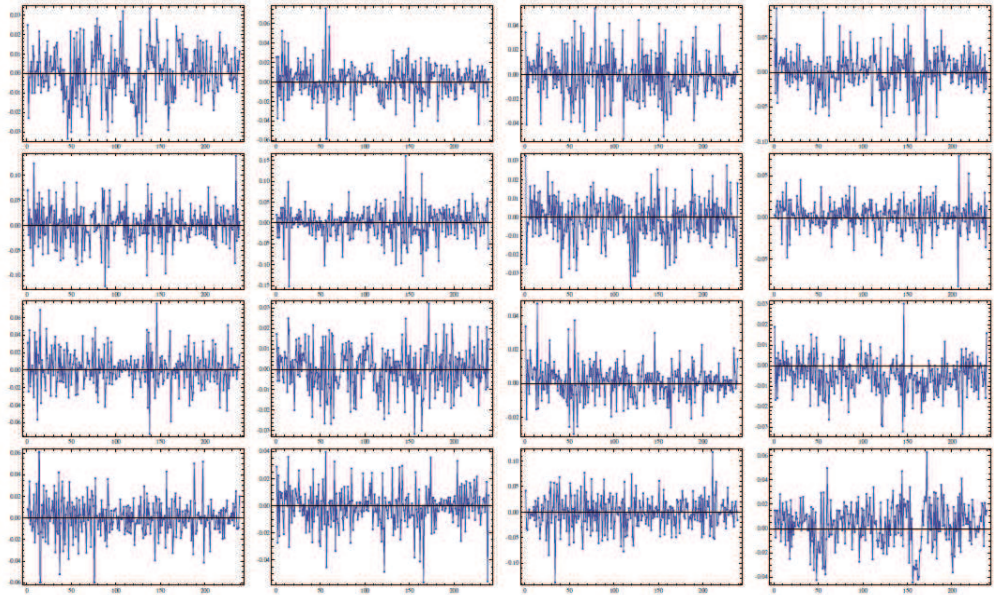}
\caption{
From the top left to the bottom right in a transverse direction, the time-series of log-returns of the indices are shown for the 16 sectors;
steel industry, non-ferrous metal industry, metal products industry, machinery industry, 
transportation equipment industry, precision machinery industry, ceramic products industry, 
chemical industry, petroleum and coal products industry, plastic products industry, 
paper and pulp industry, textile industry, food industry, other industries, 
mining industry, and electric machinery industry. 
}
\label{fig:IIP_logdiff}
\end{center}
\end{figure}

The stationarization is the important reprocessing step of the time-series analysis.
First, log-returns of the indices were calculated for the 16 industrial sectors.
Stationarity of these time-series were then confirmed using the unit root test.
The time-series of log-returns of the indices are shown for the 16 sectors in Fig.~\ref{fig:IIP_logdiff}

Fourier series expansion of the time-series of log-returns of the indices were then calculated, and some high and low frequency components were removed. The remained Fourier components were described later.

\subsection{Hilbert Transformation}\label{sec:emp:Hilbert}

The Hilbert transformation of a continuous time-series $x(t)$ is defined by Eq.(\ref{eq:Hilbert1}) \cite{Gabor1946},
\begin{equation}
 \label{eq:Hilbert1}
 y(t) = H[x(t)] = \frac{1}{\pi} PV \int_{-\infty}^\infty \frac{x(s)}{t-s} ds,
\end{equation}
where PV represents the Cauchy principal value.
A complex time-series is obtained by adopting the time-series $y(t)$ as an imaginary part.
Consequently, a phase time-series $\theta(t)$ is obtained using Eq.(\ref{eq:Hilbert2}),
\begin{equation}
 \label{eq:Hilbert2}
 z(t) = x(t) + y(t) = A(t) \exp[i \theta(t)].
\end{equation}
The following example may help the readers to understand the idea of the Hilbert transformation.
Suppose, the time-series $x(t)$ is a cosine function $x(t)=\cos(\omega t)$, 
then the Hilbert transformation of $x(t)$ will be $y(t)=H[\cos(\omega t)]=\sin(\omega t)$.
Similarly, for a sine function  $x(t)=\sin(\omega t)$, the Hilbert transformation will be $y(t)=H[\sin(\omega t)]=-\cos(\omega t)$.
Using the Euler's formula $z(t)=\cos(\omega t)+\sin(\omega t)=A(t) \exp[i \theta(t)]$, 
we can calculate the phase time-series $\theta(t)$.  

\begin{figure}
\begin{center}
\includegraphics[width=0.95\textwidth]{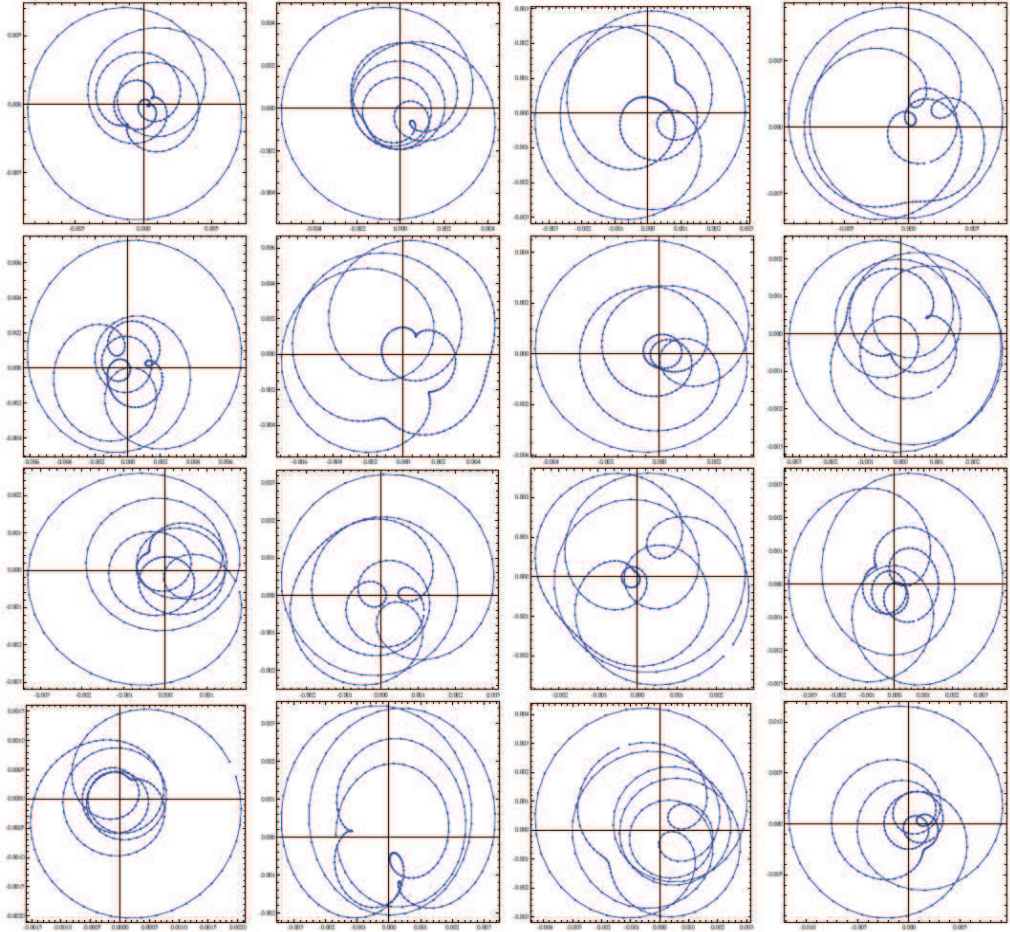}
\caption{
From the top left to the bottom right in a transverse direction, the trajectories of the time-series are shown in the complex plane, where the horizontal axis $x$ and the vertical axis $y=H[x]$, for the 16 sectors;
steel industry, non-ferrous metal industry, metal products industry, machinery industry, 
transportation equipment industry, precision machinery industry, ceramic products industry, 
chemical industry, petroleum and coal products industry, plastic products industry, 
paper and pulp industry, textile industry, food industry, other industries, 
mining industry, and electric machinery industry. 
Fourier components of oscillation period from 24 month to 80 month were included in these plots.
}
\label{fig:IIP_polar}
\end{center}
\end{figure}

In Fig.~\ref{fig:IIP_polar}, 
the trajectories of the time-series are shown in the complex plane, where log-returns of the indices $x$ and the Hilbert transformation $y=H[x]$ are used as the horizontal axis and the vertical axis, respectively, for the 16 sectors.
Here, log-returns of the indices $x$ are expanded as Fourier time-series, shown in Eq.(\ref{eq:Hilbert3}),
\begin{equation}
 \label{eq:Hilbert3}
 x(t) = \frac{A_0}{2} + \sum_{n=1}^\infty \left( A_n \cos \frac{n \pi t}{T} + B_n \sin \frac{n \pi t}{T} \right).
\end{equation}
Then the time-series $y$ is calculated using the Fourier coefficient in Eq.(\ref{eq:Hilbert3}).
\begin{align}
 \label{eq:Hilbert4}
 y(t) &= \frac{A_0}{2} + \sum_{n=1}^\infty \left( A_n H \Bigl[ \cos \frac{n \pi t}{T} \Bigr] + B_n H \Bigl[ \sin \frac{n \pi t}{T} \Bigr] \right) \notag \\
 &= \frac{A_0}{2} + \sum_{n=1}^\infty \left( A_n \sin \frac{n \pi t}{T} - B_n \cos \frac{n \pi t}{T} \right).
\end{align}
Fourier components of oscillation period from 24 month to 80 month were included in graphs of Fig.~\ref{fig:IIP_polar}.
Some irregular rotational movement was observed due to the non-periodic nature of the business cycle.

\begin{figure}
\begin{center}
\includegraphics[width=0.95\textwidth]{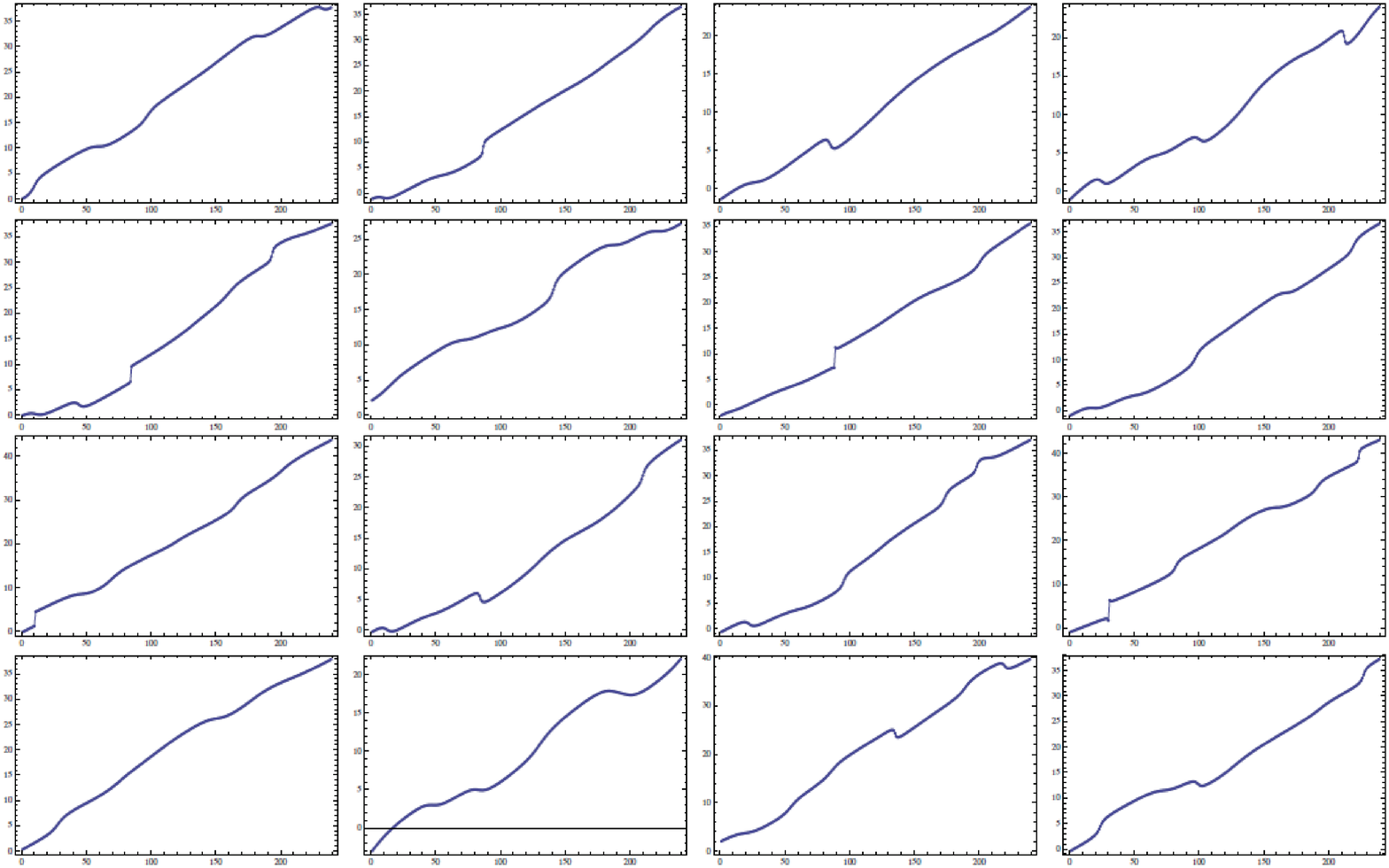}
\caption{
From the top left to the bottom right in a transverse direction, the time-series of the phase $\theta(t)$ are shown for the 16 sectors;
steel industry, non-ferrous metal industry, metal products industry, machinery industry, 
transportation equipment industry, precision machinery industry, ceramic products industry, 
chemical industry, petroleum and coal products industry, plastic products industry, 
paper and pulp industry, textile industry, food industry, other industries, 
mining industry, and electric machinery industry. 
Fourier components of oscillation period from 24 month to 80 month were included in these plots.
}
\label{fig:IIP_phase}
\end{center}
\end{figure}

The time-series of the phase $\theta (t)$ were obtained using Eq.(\ref{eq:Hilbert2}) for those 16 sectors.
Figure \ref{fig:IIP_phase} depict the time-series of the phase $\theta (t)$ for the 16 sectors.
Fourier components of oscillation period from 24 month to 80 month were included in these plots.
We observed the linear trend of rotational movement with some fluctuation for 16 sectors.
The small jumps of phases in Fig.~\ref{fig:IIP_phase} were caused by the irregular rotational movement, especially trajectories passed near the origin of the plane, which is observed in Fig.~\ref{fig:IIP_polar}.

\section{Discussion}\label{sec:emp:Shock}

\subsection{Frequency Entrainment}\label{sec:discuss:Frequency}

\begin{figure}
\begin{center}
\includegraphics[width=0.45\textwidth]{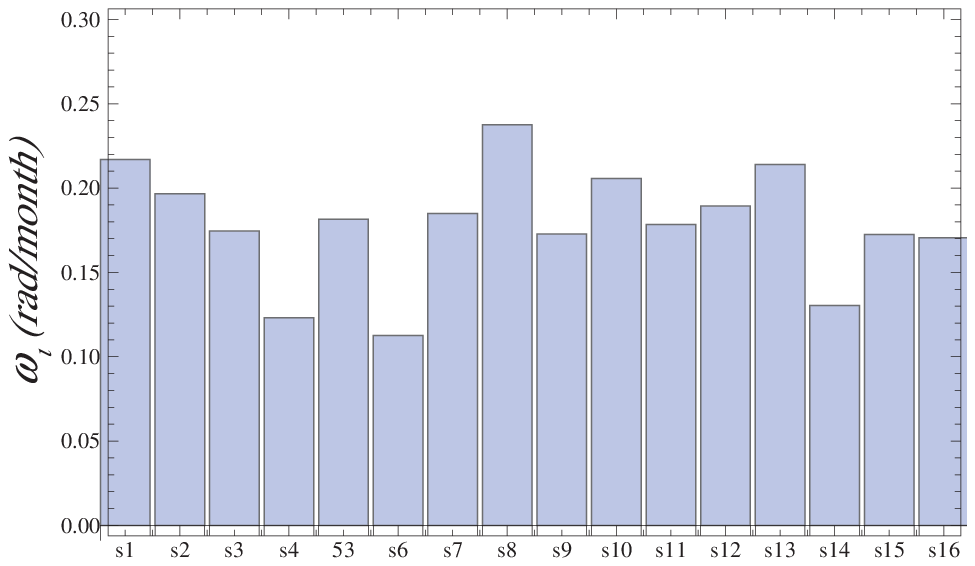}
\includegraphics[width=0.45\textwidth]{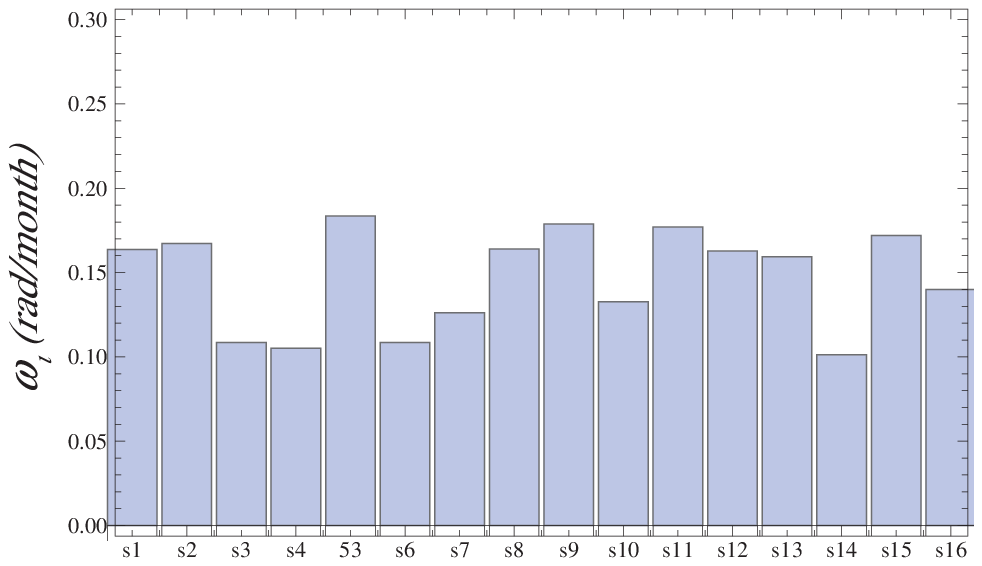}
\caption{
The estimated angular frequencies $\omega_i$ are shown for the Fourier components: 18 to 80 months (left) and  24 to 80 months (right).
}
\label{fig:AF}
\end{center}
\end{figure}

The frequency entrainment and phase locking are expected to be observed as the direct evidence of the synchronization.
The angular frequency $\omega_i$ and the intercept $\tilde{\theta}_i$ are estimated by fitting the time-series of the phase $\theta_i(t)$
using Eq.(\ref{eq:Freq1}).
\begin{equation}
 \label{eq:Freq1}
 \theta_i(t) = \omega_i t + \tilde{\theta}_i,
\end{equation}
where $i$ indicates the industrial sector.
The estimated angular frequencies $\omega_i$ for all the 16 sectors are plotted for the Fourier components: 
18 to 80 months and  24 to 80 months in Fig.~\ref{fig:AF}, where we observed that the estimated angular frequencies $\omega_i$ are entrained to be almost identical for the 16 sectors, 
and this means that the frequency entrainment is observed.
When the higher frequency Fourier components are included in the analysis, 
the deviation of the estimated angular frequencies $\omega_i$ among the 16 sectors are expected to increase gradually.
The increase in the deviation can be explained from the presence of larger jumps of phases due to the trajectories passing near the origin of the complex plane for the higher frequency Fourier components.

\subsection{Phase Locking}\label{sec:discuss:Phase}

\begin{figure}
\begin{center}
\includegraphics[width=0.45\textwidth]{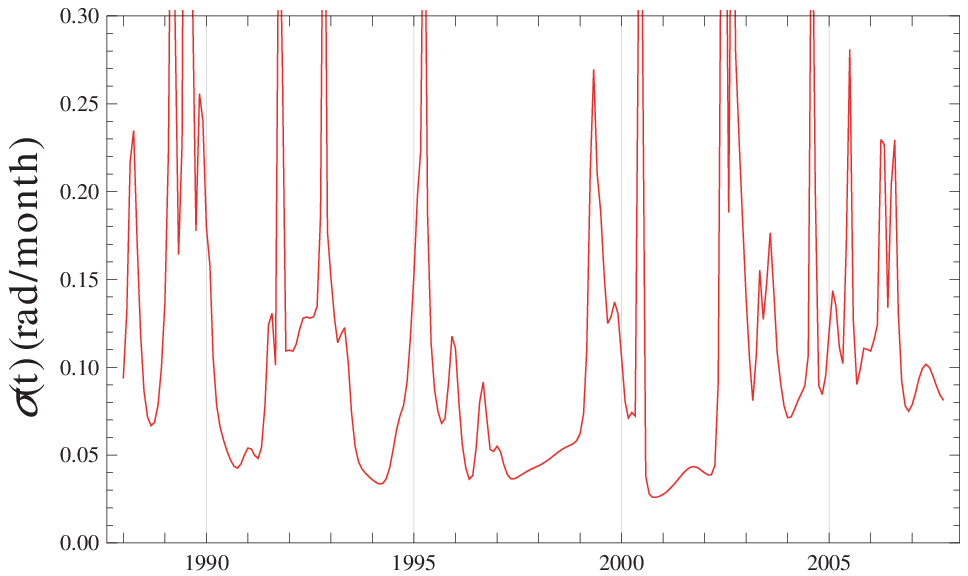}
\includegraphics[width=0.45\textwidth]{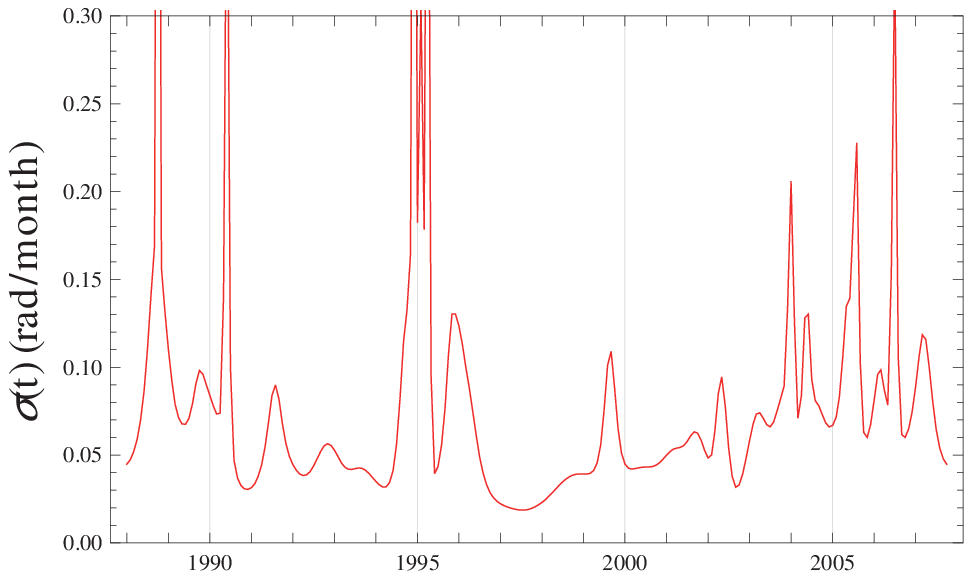}
\caption{
The estimated indicators of the phase locking $\sigma (t)$ are plotted for the Fourier components: 18 to 80 months (left) and  24 to 80 months (right).
}
\label{fig:PhaseLock}
\end{center}
\end{figure}

Phase locking means that phase differences for all pairs of oscillators are constant. 
However, this condition is rarely seen to exactly satisfy the actual time-series due to irregular fluctuation, i.e., economic shock.
We introduce an indicator $\sigma (t)$ of the phase locking as
\begin{equation}
 \label{eq:PhaseLock1}
 \sigma(t) = \Bigl[ \frac{1}{16} \sum_{i=1}^{16} \Bigl\{ \frac{d}{dt}(\theta_i(t)-\omega_i t) - \mu(t) \Bigr\}^2 \Bigr]^{1/2},
\end{equation}
\begin{equation}
 \label{eq:PhaseLock2}
 \mu(t) = \frac{1}{16} \sum_{i=1}^{16} \frac{d}{dt}(\theta_i(t)-\omega_i t).
\end{equation}
The indicator $\sigma (t)$ is equal to zero when the phase differences for all pairs of oscillators are constant. 
On the other hand, if the indicator $\sigma (t)$ satisfies the following relation, then we call this the partial phase locking.
\begin{equation}
 \label{eq:PhaseLock3}
 \sigma(t) \ll \omega_i, 
\end{equation}
The estimated indicator of the phase locking $\sigma (t)$ is plotted in Fig.~\ref{fig:PhaseLock} for the Fourier components: 18 to 80 months and  24 to 80 months. 
Figure \ref{fig:PhaseLock} depicts that the indicator $\sigma (t)$ is much smaller than $\omega_i$ for most of the period. 
This means that the partial phase locking is observed.
As a consequence of this analysis, both the frequency entrainment and the phase locking are obtained as the direct evidence of the synchronization.

\subsection{Common Shock and Individual Shock}\label{sec:discuss:Shock}

\begin{figure}
\begin{center}
\includegraphics[width=0.45\textwidth]{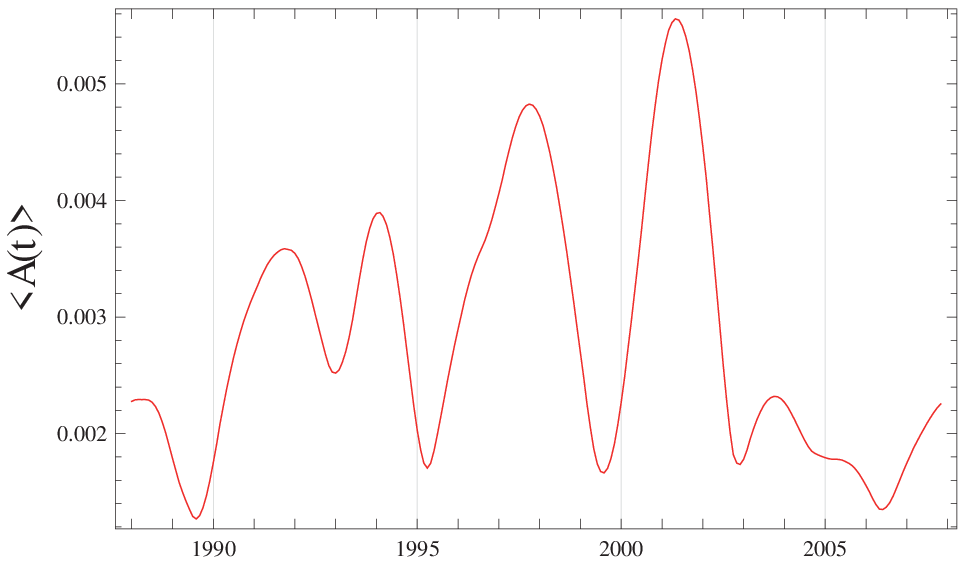}
\includegraphics[width=0.45\textwidth]{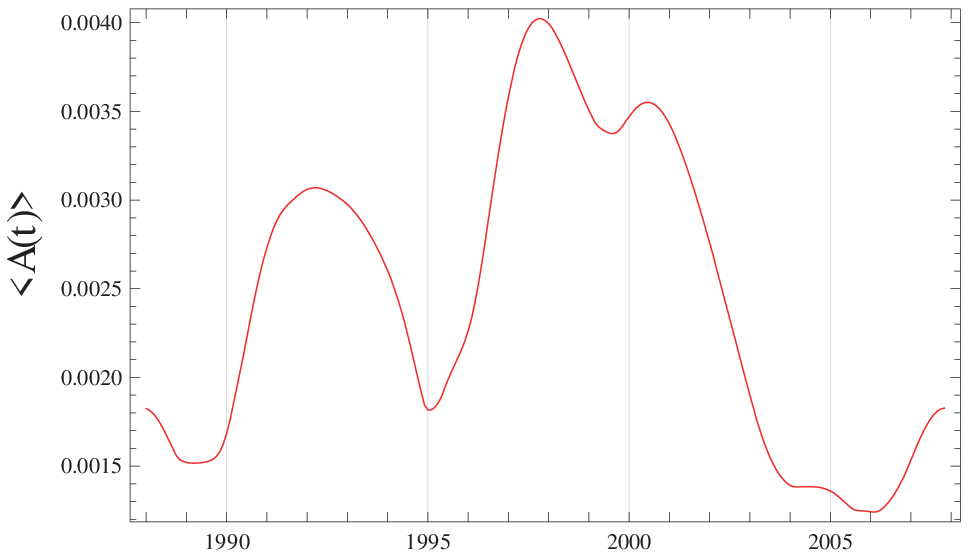}
\caption{
The average amplitudes $\langle A(t) \rangle$ are shown for the Fourier components: 18 to 80 months (left) and  24 to 80 months (right).
}
\label{fig:Amplitude}
\end{center}
\end{figure}

\begin{figure}
\begin{center}
\includegraphics[width=0.45\textwidth]{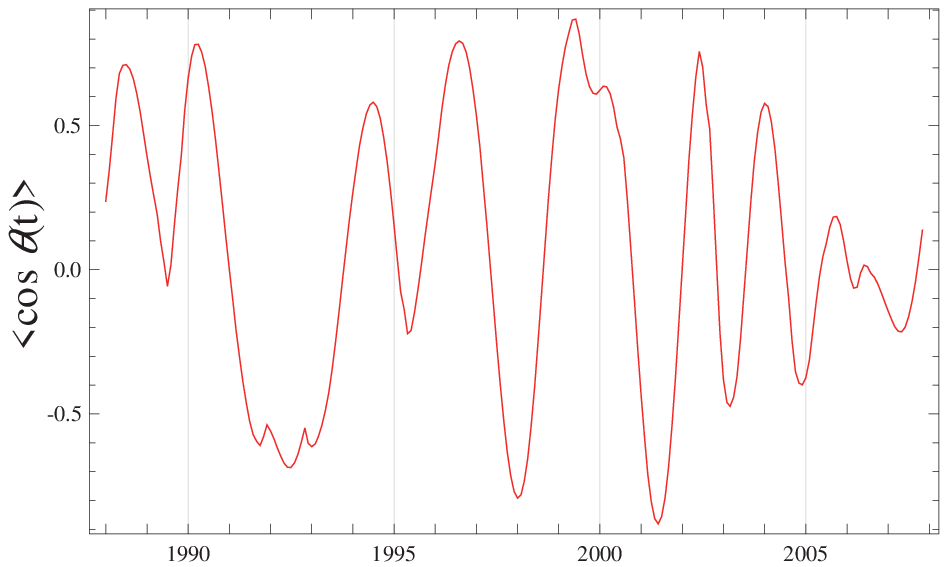}
\includegraphics[width=0.45\textwidth]{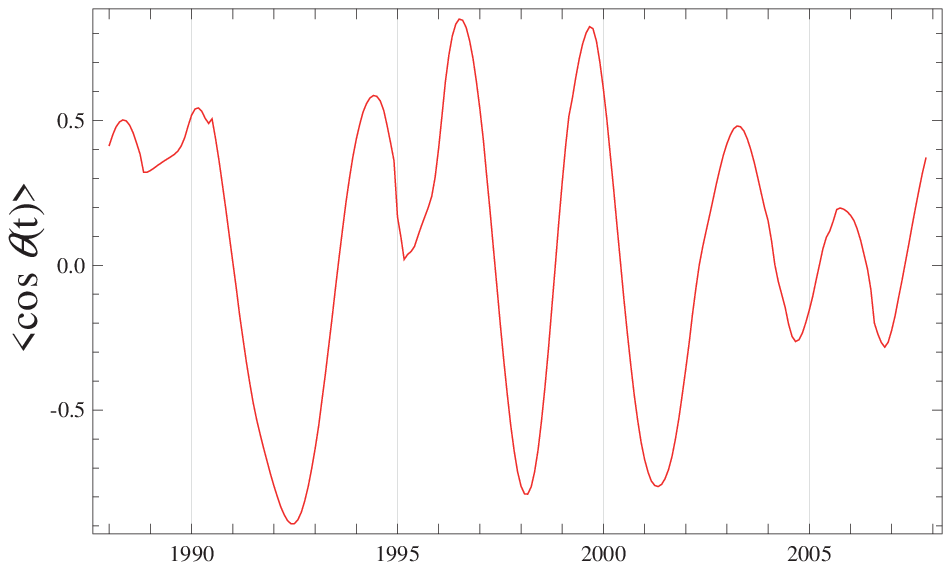}
\caption{
The average phases $\langle \cos \theta(t) \rangle$ are shown for the Fourier components: 18 to 80 months (left) and  24 to 80 months (right).
The average phases $\langle \cos \theta(t) \rangle$ show sharp drop in all of 1992, 1998 and 2001.
It should be noted that the phase $\theta_i (t)$ is more sensitive to economic shock than the amplitude $A_i (t)$.
}
\label{fig:CommonShock}
\end{center}
\end{figure}

The log-returns of the indices $x_i$ are decomposed to the amplitude $A_i (t)$ and the phase $\theta_i (t)$ using Eq.(\ref{eq:Hilbert1}) and  Eq.(\ref{eq:Hilbert2}). It is interesting to ask the question ``Which quantity carry the information about the economic shock, the amplitude $A_i (t)$ or the phase $\theta_i (t)$?''.
The averages of these quantities over the industrial sectors are written as:
\begin{equation}
 \label{eq:Shock1}
 \langle A(t) \rangle = \frac{1}{16} \sum_{i=1}^{16} A_i(t) = \frac{1}{16} \sum_{i=1}^{16} \frac{x_i(t)}{\cos \theta_i(t)}, 
\end{equation}
\begin{equation}
 \label{eq:Shock2}
 \langle \cos \theta(t) \rangle = \frac{1}{16} \sum_{i=1}^{16} \cos \theta_i(t).
\end{equation}
The average amplitudes $\langle A(t) \rangle$ and the average phases $\langle \cos \theta(t) \rangle$ are shown in Fig.~\ref{fig:Amplitude} and Fig.~\ref{fig:CommonShock}, respectively.
These figures clearly show that the information about the economic shock is carried by the phase $\theta_i (t)$, not by the amplitude $A_i (t)$.
In Japan, we had severe economic recessions in 1992, 1998, and 2001. The recession in 1992 was due to a collapse of bubble economy.
In 1998, we had a banking crisis, when actually some major banks went bankrupt.
The recession in 2001 was due to bursting of the information technology bubble.
The average amplitudes $\langle A(t) \rangle$ in Fig.~\ref{fig:Amplitude} have large values in all of 1992, 1998 and 2001.
On the contrary,the average phases $\langle \cos \theta(t) \rangle$ in Fig.~\ref{fig:CommonShock} show sharp drop in all of 1992, 1998, and 2001.
It should be noted that the phase $\theta_i (t)$ is more sensitive to economic shock than the amplitude $A_i (t)$.
Thus it is adequate to interpret the average phases $\langle \cos \theta(t) \rangle$ and the residual $\theta_i (t) - \langle \cos \theta(t) \rangle$ as common shock and individual shock, respectively.

\begin{figure}
\begin{center}
\includegraphics[width=0.45\textwidth]{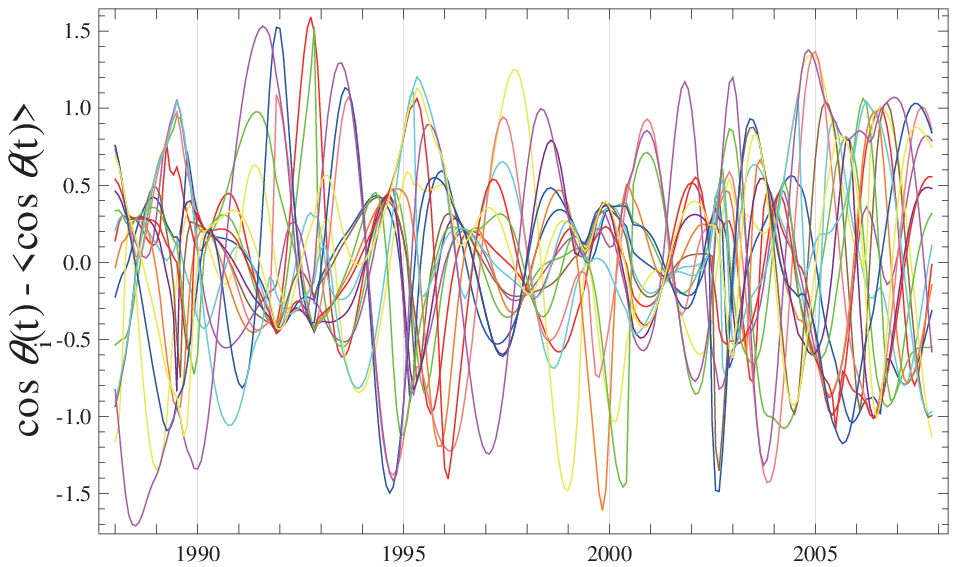}
\includegraphics[width=0.45\textwidth]{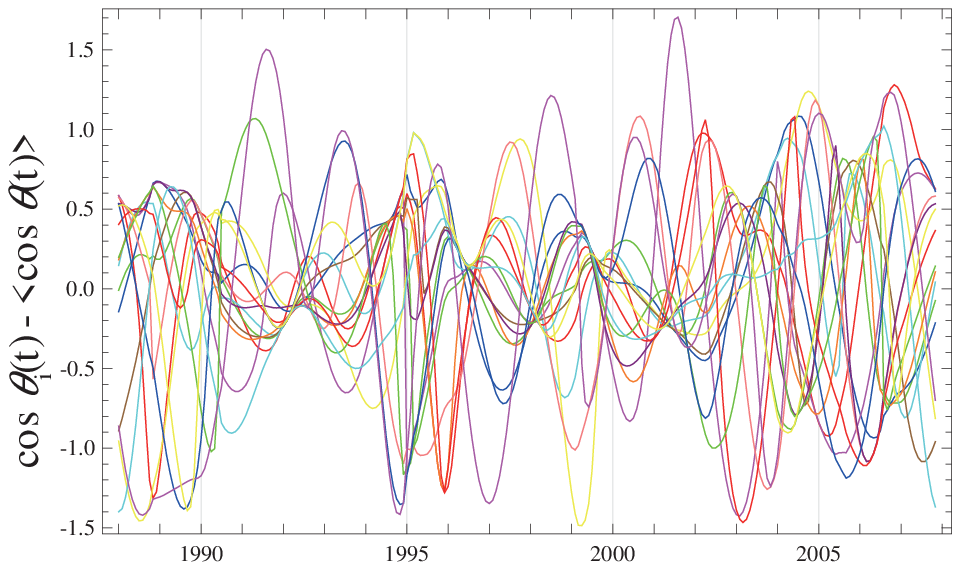}
\caption{
The individual shocks are shown for the Fourier components: 18 to 80 months (left) and 24 to 80 months (right).
}
\label{fig:IndividualShock}
\end{center}
\end{figure}

The individual shocks $\cos \theta_i(t) - \langle \cos \theta(t) \rangle$ for each industrial sector are shown in Fig.~\ref{fig:IndividualShock}.
It depicts that we have many individual shocks all the time, and many of them seem to occur randomly. 
More noteworthy are contraction of the individual shocks in 1992, 1998 and 2001, when we had severe economic recessions.
All industries were exposed to the common shocks $\langle \cos \theta(t) \rangle$ during this periods.

\section{Summary}\label{sec:summary}

We analyzed the Indices of Industrial Production (Seasonal Adjustment Index) for a long period of 240 months (January 1988 to December 2007) to develop a deeper understanding of the economic shocks. 
The angular frequencies $\omega_i$ estimated using the Hilbert transformation are almost identical for the 16 sectors.
Moreover, indicator of the phase locking $\sigma (t)$ shows that the partial phase locking is observed for the 16 sectors.
These are the direct evidence of the synchronization in the Japanese business cycle.
In addition to this, we also showed that the information of the economic shock is carried by the phase time-series. The common shock and individual shocks are separated using phase time-series $\theta_i(t)$. 
It is interpreted the average phases $\langle \cos \theta(t) \rangle$ and the residual $\theta_i (t) - \langle \cos \theta(t) \rangle$ as common shock and individual shock, respectively.
The former dominates the economic shock in all of 1992, 1998 and 2001. In the same periods, the individual shock has contraction.
We have many individual shocks all the time, and individual shocks seem to occur randomly. 
The obtained results suggest that the business cycle may be described as a dynamics of the coupled limit-cycle oscillators exposed to the common shocks and random individual shocks.

\section*{Acknowledgements}
We thank Yoshi Fujiwara and Wataru Souma for valuable discussion and comments. This work is
supported in part by the Program for Promoting Methodological
Innovation in Humanities and Social Sciences by Cross-Disciplinary
Fusing of the Japan Society for the Promotion of Science.






\begin{thebibliography}{99}
\bibitem{Mitchell1964}
A.~F.~Burns and W.~ C.~Mitchell., ``Measuring Business Cycles", National Bureau of Economic Research (Studies in business cycles; 2), 1964. 
\bibitem{Yoshikawa2000}
H.~Yoshikawa, ``Macroeconomics", Chapter 2, Tokyo: Sobunsha, 2000. (In Japanese.)
\bibitem{Aoki2007}
M.~Aoki and H.~Yoshikawa, ``Reconstructing Macroeconomic - A Perspective from Statistical Physics and Combinatorial Stochastic Processes", Cambridge: Cambridge University Press, 2007.
\bibitem{Granger1964}
C.W.J. Granger and M. Hatanaka, ``Spectral Analysis of Economic Time Series" , Princeton: Princeton University Press,1964.
\bibitem{Samuelson1939}
P.~Samuelson, ``Interactions bwteen the Multiplier Analysis and the Principle of Accelerator", Review of Economic Statistics, May 1939.
\bibitem{Hickes1950}
J.~Hickes, ``A Contribution to the Theory of the Trade Cycle", Oxford: Oxford University Press, 1950.
\bibitem{Goodwin1951}
R.~Goodwin, ``The Nonlinear Accelerator and the Persistence of Business Cycles", Econometrica, January 1951.
\bibitem{Kaldor1940}
N.~Kaldor, ``A Model of the Trade Cycle", Economic Journal 50, pp.78-92, 1940
\bibitem{Lorenz1989}
H.W.~Lorenz, ``Nonlinear Dynamical Economics and Chaotic Motion", Chapter 2 and 5, Springer-Verlag Berlin Heidelberg, 1989
\bibitem{Moseklide1992}
E.~Moseklide, E.R.~Larsen, J.D.~Sterman, and J.S.~Thomsen, ``Non-Linear Model-Interaction in the Macroeconomy", Annals of Operation Research 37, pp.185-215, 1992. 
\bibitem{Sterman1994}
J.D.~Sterman, and E.~Moseklide, ``Business Cycles and Long Waves: A Behavioral Disequilibrium Perspective", W.~Semmler (ed.): Business Cycles: Theory and Empirical Methods, Dordrecht: Kluwer, 1994.
\bibitem{Tinbergen1939}
J.~Tinbergen, ``Business cycles in the United States of America, 1919-1932",  -- League of Nations, Economic Intelligence Service, 1939. 
\bibitem{Kydland1982}
F.~Kydland and E.~Prescott, ``Time to Build and Aggregate Flucuations", Econometrica 50,pp.1354-1370, 1982.
\bibitem{Slutsky1937}
E.~Slutsky, ``The Summation of Random Causes as a Source of Cyclical Processes", Econometrics 5, pp.105-146, 1937.
\bibitem{Huygens1673}
C.~Huygens, ``Horologium oscillatorium: 1673", Michigan: Dawson, 1966.
\bibitem{Kuramoto1984}
Y.~Kuramoto, ``Chemical Sccillations, Waves, and Turbulence", Springer-Verlag Berlin Heidelberg, 1984.
\bibitem{Ikeda2012}
Y.~Ikeda, H.~Aoyama, Y.~Fujiwara, H.~Iyetomi, K.~Ogimoto, W.~Souma, and H.~Yoshikawa, ``Coupled Oscillator Model of the Business Cycle with Flucuating Goods Market", Progress of Theoretical Physics Supplement 194, pp.111-121, 2012.
\bibitem{Gabor1946}
D.~Gabor, ``Theory of communication. Part 1: The analysis of information", J. Inst. Elect. Eng. 93, pp.429-441, 1946.
\end{thebibliography}

\end{document}